\documentclass[12pt,preprint]{aastex61}
\renewcommand{\baselinestretch}{1.1}
\usepackage{graphicx}
\usepackage{amsmath,amssymb,latexsym}
\usepackage{multirow}
\usepackage{natbib}
\usepackage{url}
\usepackage{graphicx}

\newcommand{\sitem}{\vspace{-0.65cm}\item}
\shorttitle{Jupiter and Saturn Interiors}
\shortauthors{Helled, R.}

\begin{document}

\title{The Interiors of Jupiter and Saturn}
\author{Ravit Helled}
\affil{
Center for Theoretical Astrophysics \& Cosmology, 
Institute for Computational Science, \\
University of Zurich, Switzerland.\\ }

\begin{abstract}
Probing the interiors of the gas giant planets in our Solar System is not an easy task. It requires a set of accurate measurements combined with theoretical models that are used to infer the planetary composition and its depth dependence. 
The masses of Jupiter and Saturn are 317.83 and 95.16 Earth masses (M$_{\oplus}$), respectively, and since a few decades, we know that they mostly consist of hydrogen and helium.Ê
It is the mass of heavy elements (all elements heavier than helium) that is not well determined, as well as their distribution within the planets.Ê
While the heavy elements are not the dominating materials in Jupiter and Saturn they are the key for our understanding of their formation and evolution histories.Ê

The planetary internal structure is inferred from theoretical models that fit the available observational constraints by using theoretical equations of states (EOSs) for hydrogen, helium, their mixtures, and heavier elements (typically rocks and/or ices). Ê
However, there is no unique solution for the planetary structure and the results depend on the used EOSs and the model assumptions imposed by the modeller.Ê

Major model assumptions that can affect the derived internal structure include the number of layers, the heat transport mechanism within the planet (and its entropy), the nature of the core (compact vs. diluted), and the location (pressure) of separation between the two envelopes. 
Alternative structure models assume a less distinct division between the layers and/or a non-homogenous distribution of the heavy elements. The fact that the behaviour of hydrogen at high pressures and temperatures is not perfectly known, and that helium may separate from hydrogen at the deep interior add sources of uncertainties to structure models. Today, with accurate measurements of the gravitational fields of Jupiter and Saturn from the Juno and Cassini missions, structure models can be further constrained. At the same time, these measurements introduce new challenges for planetary modellers. 
\end{abstract}


\section{Introduction}
Investigating the interiors of the giant planets in the Solar System goes back several decades. 
Jupiter and Saturn are located at radial distances of about 5.2 and 9.6 AU\footnote{1AU is an astronomical unit, the average distance between the Earth and the Sun.} from the Sun and their composition is dominated by light elements, in particular, hydrogen and helium (hereafter H-He). Jupiter and Saturn are massive, fast rotators, and their atmospheres are characterised by impressive signatures of dynamics. 
These colourful atmospheres, however, represent only the "skin" of the planets, and cannot reveal the secrets of their internal structures. 
Therefore despite the significant progress on both the observational and theoretical fronts, Jupiter and Saturn remain mysterious planets.  
\par

Due to their large distances from the Earth and their gaseous nature, revealing information on the deep interiors of Jupiter and Saturn must be done by using indirect measurements. 
As more information about the planets is collected, more comprehensive theoretical structure models must be developed. 
However, it becomes increasingly challenging to find a  self-consistent theoretical framework that meets all the observational constraints. 
The accurate measurements of the physical and chemical planetary properties provide new and important constraints, but they also lead to new open questions. 
The current situation in modelling planetary interiors can be summarised by Albert Einstein's quote "{\it The more I learn, the more I realise how much I don't know}". 
Fitting the new data requires more complex structure models and the inclusion of various physical processes and assumptions which are not well-justified and/or completely understood. 
\par 

Currently, we still do not have unique and self-consistent views of the interiors of Jupiter and Saturn, 
but we have a better understanding of the relevant physical/chemical processes that should be considered, and of the limitation of our theoretical approaches. 
Making progress in that direction, does not only help us to study planetary interiors but also to better understand the behaviour of simple elements at high pressures and temperatures, and to put important constraints on giant planet formation and evolution models.  
\par

In this review, I summarise the current knowledge of the internal structures of Jupiter and Saturn. 
Recent reviews of topic include Militzer et al.~, 2016, Guillot \& Gautier, 2016, Helled \& Guillot, 2018;  Baraffe et al.,~2014, Fortney \& Nettelmann, 2010, Fortney et al. 2017 and references therein. 

\section{Making an Interior Model}

\subsection{Observational Constraints}
Internal structure models are designed to fit the observed physical data of the planets, such as their masses, radii, gravitational and magnetic fields, 1-bar temperatures, atmospheric composition, and internal rotations. Key physical properties of Jupiter and Saturn  are summarised in Table 1. 
Interestingly, the atmospheres of both Jupiter and Saturn show a depletion in helium in comparison to the helium mass fraction of the proto-solar value of $Y_{proto} \sim$ 0.275 as inferred from stellar evolution models for the Sun (e.g. Bahcall et al. 1995). 
The measured helium mass fractions in Jupiter and Saturn are found to be $\sim 0.238$ (von Zahn et al. 1998) and 0.18--0.25 (Conrath \& Gautier 2000), respectively. 
As discussed below, this does not imply that Jupiter and Saturn are depleted in helium globally but instead, that the distribution of helium is inhomogeneous within their interiors due to the phenomenon of helium settling (see section 2.3.2 for details). 
In addition, for Jupiter, the Galileo entry probe provided abundances measurements of other components suggesting that Jupiter's outer envelope is enriched with heavy elements by a factor of $\sim 2-4$ compared to proto-solar abundance (e.g., Atreya et al., 2003; Guillot \& Gautier, 2014). 
Two exceptions were neon and oxygen, both found to be depleted, however neon is expected to be affected by the process of helium rain (Roustlon \& Stevenson, 1995, Wilson \& Militzer, 2010)  and the low 
abundance of water is probably linked to the special location of the probe's entry which is known as a "dry spot" where the atmosphere is dry and does not represent the bulk of the atmosphere. Therefore, at the moment the oxygen abundance in Jupiter is still unknown.  

Since giant planets consist of mostly fluid hydrogen and helium, they do not have a solid surface below the clouds like terrestrial planets. 
Therefore, the "surface" of the planet is defined as the location where the pressure is 1 bar, comparable to the pressure at the Earth's surface. 
The temperature at this location is measured (with a small uncertainty). Then, with information of the temperature at 1 bar, the entropy of the outer envelope is determined and adiabatic models\footnote{In adiabatic models the temperature profile is set by the adiabatic gradient, and the entropy is (nearly) constant within the planet (see Militzer et al., 2016 and references therein for details).} can be constructed. 

The gravitational field of the planet, or more precisely, the total potential which also includes the rotational term and is given by, 
 \begin{eqnarray}
U &=& \frac{G M}{r} \left( 1 - \sum_{n=1}^\infty \left( \frac{a}{r} \right)^{n} \mathrm{J}_{n} \mathrm{P}_{n} \left( \cos \theta  \right)  \right) + \frac{1}{2} \omega^2 r^2 \sin^2 \theta .
\label{U}
\end{eqnarray}
where (r, $\theta, \phi$) are spherical polar coordinates, $a$ is the equatorial radius, and $M$ is the total planetary mass. 
The potential $U$ is represented as an expansion in Legendre polynomials (e.g., Zharkov and Trubitsyn, 1978), where typically only the even indices (i.e., P$_{2 n}$) are taken into account due to the
primarily north/south symmetry of the two hemispheres\footnote{This is however a simplification as odd harmonics have been measured for Jupiter with the Juno spacecraft (Iess et al., 2018). 
The measurement of Jupiter's gravitational field being north--south asymmetric, has been used to reveal the planetÕs atmospheric and interior flows (Kaspi et al., 2018).}. 
The gravitational harmonic coefficients J$_{2 n}$ are 
typically inferred from Doppler tracking data of a spacecraft orbiting or flying by the planet and are used to constrain the density profile as discussed below.

\begin{table}[h!]
\def\arraystretch{1.}
\centering
\label{tab:1}   
\begin{tabular}{p{4.5cm}p{3.2cm}p{3.2cm}}
\hline\noalign{\smallskip}
 {\bf Physical Property} & {\bf Jupiter} & {\bf Saturn }  \\
 \hline
Distance to Sun (AU) & 5.204 & 9.582\\
Mass (10$^{24}$ kg)& 1898.13$\pm$0.19 &  568.319$\pm$0.057\\
Mean Radius (km) & 69911$\pm$6& 58232$\pm$6 \\
Equatorial Radius (km) & 71492$\pm$4& 60268$\pm$4 \\
Mean Density (g/cm$^{3}$) & 1.3262$\pm$0.0004 & 0.6871$\pm$0.0002 \\
$J_2 \times 10^6$ & 14696.572$\pm$0.014 & 16290.7$\pm$0.27  \\
$J_4 \times 10^6$ & -586.609$\pm$0.004  & -935.83$\pm$2.77  \\
$J_6 \times 10^6$ & 34.24$\pm$0.24 & 86.14$\pm$9.64 \\
$C/MR_{eq}^2$ (MOI)$^i$  & 0.264  & 0.22 \\
Rotation Period  & 9h 55m 29.56s &  10h 39m $\pm\sim$10m\\
Effective Temperature (K) & 124.4$\pm$0.3 & 95.0$\pm$0.4\\
1-bar Temperature (K) & 165$\pm$4 & 135$\pm$5 \\
\noalign{\smallskip}\hline\noalign{\smallskip}
\end{tabular}
\caption{Basic observed properties of Jupiter and Saturn from Helled \& Guillot (2018), https://ssd.jpl.nasa.gov/ and references therein. 
Jupiter's gravitational field is taken from Iess et al.~(2018). The gravitational coefficients correspond to the reference equatorial radii of 71,492 km and 60,330 km for Jupiter and Saturn, respectively (http://ssd.jpl.nasa.gov/?gravity\_fields\_op). 
\\
\footnotesize{$^i$these are theoretical values based on interior model calculations. $^{ii}$see section 4.1 and Helled et al.~2015 for discussion on Saturn's rotation rate uncertainty.}
}
\end{table}

\subsection{Governing Equations}
The planetary interior is modelled by solving the standard structure equations  which include the mass conservation, hydrostatic balance, and thermodynamic equations as follows:
\begin{equation}
\frac{dm}{dr} = 4\pi r^2\rho,
\end{equation}
\begin{equation}
\frac{1}{\rho}\frac{dP}{dr}=-{Gm\over r^2} + \frac{2}{3}\omega^2 r
\end{equation}
\begin{equation}
{dT\over dr}=\frac{T}{P}\frac{dP}{dr}\nabla_T,
\end{equation}
where $P$ is the pressure, $\rho$ is the density, $m$ is the mass within a sphere of a radius $r$, and $\omega$ is the rotation rate. 
In order to account for rotation, the hydrostatic equation (Eq.~2) includes an additional term which depends on $\omega$, which is assumed to be constant (i.e., uniform rotation), for a non-spinning planet, $\omega=0$. 
For a rapidly rotating planet, this equation is valid in the limit of a barotropic fluid and a solid-body rotation. The radius $r$ is then considered as a mean volumetric radius. Eq. (3) is the first order expansion of the total potential $U$. 
The temperature gradient $\nabla_T\equiv d\ln T /d\ln P$ depends on the heat transport mechanism (convection vs.~conduction/radiation).
Typically, the temperature gradient is taken to be the smallest among the adiabatic $\nabla_{ad}$, radiative/conductive $\nabla_{rad/cond}$ gradients since the heat transport mechanism that leads to the smallest temperature gradient is the most efficient one. 
In other words, the temperature gradient is taken to be $\nabla_T = min[\nabla_{ad}, \nabla_{rad/cond}]
$\footnote{The adiabatic gradient $\nabla_{ad}$=$\frac{\partial lnT}{\partial lnP}\arrowvert_s$, where $S$ is the entropy, corresponds to a case in which the material is homogenous and convective. 
The radiative/conductive gradient is given by $\nabla_{rad/cond} = \frac{3 \kappa L P }{64 \pi \sigma T^4 G m}$, where $\kappa$ is the Rosseland opacity which accounts for contributions from both radiation and conduction, and $\sigma$ is the Stephan-Boltzmann constant (see Guillot et al.~(2004), Militzer et al. (2016) and references therein for further details).}.
Finally, in order to solve this set of equations we need to know how the density depends on the temperature and pressure, i.e., $\rho(P,T)$ which is determined by the equation of state. 
\par

The density profile of the planets is set to reproduce the measured gravitational moments J$_{2n}$. The relation between the gravitational moments and the density profile is given by (e.g., Zharkov and Trubitsyn, 1978):
\begin{equation}
Ma^{2n}\mathrm{J}_{2n} = - \int_\tau \rho(r) r^{2n} \mathrm{P}_{2n}(\cos \theta) d \tau,
\label{Jn}
\end{equation}
where the integration is carried out over the volume $\tau$. 
Traditionally, the theoretical density profile and gravitational moments were calculated by using the {\it theory of figures} (TOF) in which the harmonics are computed from a 
series approximation in the smallness parameter $m=\omega^2R^3/GM$, where $R$ is the mean radius of the planet, typically up to an order of three or four. 
This was sufficient as long as we did not have information about the gravitational coefficients beyond $J_6$ and the measurements have relatively large uncertainties.  
An alternative method to TOF that is designed to be compatible with accurate data and provide estimates for the higher order harmonics was developed by W.~Hubbard (e.g., Hubbard 2012, 2013). 
In this approach, called concentric Maclaurin spheroid (CMS), the density profile is represented by a large number of Maclaurin spheroids where a 
continuous density can be achieved if the number of spheroids is large enough. 
While the computational resources needed are large, the gravitational coefficients can be calculated to any order with an excellent precision ($\sim10^{-9}$).  
Discussion and comparison between the TOF and CMS methods can be found in Hubbard et al.~(2014), Wisdom \& Hubbard (2016), Nettelmann (2017), Debras \& Chabrier (2018) and references therein.

\subsection{Equation of State}
In thermodynamics, the equation of state (EOS) relates the state variables such as the temperature, pressure, density, internal energy, and entropy. 
Since Jupiter and Saturn are mostly composed of hydrogen and helium, modelling their structures relies on information of the EOS of hydrogen, helium, and their mixture.   
Giant planet interiors serve as natural laboratories for studying different elements at exotic conditions that do not exist on Earth. 
At the same time, calculating the EOS of materials in Jupiter and Saturn interior conditions is a challenging task because molecules, atoms, ions and electrons coexist and interact, and the pressure and temperature range varies by several orders of magnitude, 
going up to several tens of mega-bars (Mbar), i.e., 100 GPa and several 10$^4$ Kelvins. 
Therefore, information on the EOS at such conditions requires performing high-pressure experiments and/or solving the many-body quantum mechanical problem to produce theoretical EOS tables that cover such a large range of 
pressures and temperatures. 
Despite the challenges, there have been significant advances in high-pressure experiments and {\it ab initio} EOS calculations. 
Below, I briefly describe the EOS of hydrogen, helium and heavier elements. More information on that topic can be found in Fortney \& Nettelmann (2010); Baraffe et al.~(2014); Militzer et al.~(2016);  Guillot \& Gautier (2016); Helled \& Guillot (2018) and references therein. 
\par 

\subsubsection{Hydrogen}
Hydrogen is the most abundant element in the Universe, and yet, its phase diagram is still a topic of intensive research. 
The behaviour of hydrogen at high pressures can be investigated experimentally and theoretically. 
There are several types of laboratory experiments such as gas-guns, convergent shock wave, and laser-induced shock compression, that can probe hydrogen (or its isotope, deuterium) at Mbar pressures. 
Unfortunately, the available experimental data have a large range, and each experiment suffers from different limitations and systematics. 
Nevertheless, recently, some progress has been made towards convergence when comparing the shock Hugoniot curve of hydrogen from different studies (e.g., Miguel et al., 2016).  
Although empirical, the information from experiments is limited, and theoretical calculations are required to provide a wide-range EOS for hydrogen. 
On the other hand, the laboratory experiments results are of primary importance since they are used to calibrate the theoretical EOS used for planetary modelling. 
\par

The most popular {\it ab-initio} technique in materials science, and the most common approach to probe the EOS of hydrogen and helium at planetary conditions, 
is Density Functional Theory (DFT). Although the theory is exact, all existing practical implementations relies on approximations. While DFT calculations provide a 
relatively accurate determination of the EOS of hydrogen in a large range of temperatures and pressures, using moderate computational resources, 
it shows a poor performance in assessing phase transitions (e.g., Azadi \& Foulkes, 2013). 
An alternative approach is Quantum Monte Carlo (QMC), which is a wave-function-based method and can accurately solve the electronic problem (e.g., Mazzola et al., 2018 and references therein).  
This approach is much more  computationally expensive, but is potentially one-order of magnitude more accurate than DFT (e.g., Foulkes et al., 2001) and therefore can closely simulate the phase transitions, where the (free) energy difference between the competing phases (at given thermodynamic conditions) can be small. 
At the moment, QMC calculations can be used to calibrate other existing wide-range EOS tables, and are expected to play an important role in EOS calculations in the upcoming years.  
\par 

Figure 1 shows the phase diagram of dense hydrogen and hydrogen-helium mixture. Shown is the transition between the insulating-molecular and the metallic-atomic hydrogen fluid (shaded area). 
The location of this transition is not only important for understanding the generation of strong magnetic fields in these planets but also for determining the division of the planets to layers with different compositions as we discuss below. 
Also shown are Jupiter's and Saturn's adiabats as calculated by structure models (brown and mustard-colour lines, respectively).  
 \par
 There are several interesting conclusions about Jupiter and Saturn that can be made simply by looking at this the hydrogen phase diagram.  
 First, both planets lie in the regime above solid hydrogen, suggesting that they are fluid planets, as already suggested by Hubbard (1968). 
 Second, both planets cross the critical point of hydrogen which indicates that in the outer parts of the planets hydrogen is in the molecular form (H$_2$) and in the metallic form in the 
 deep interiors. 
 Metallic hydrogen is a phase of hydrogen at high pressures/temperatures where electrons are free and hydrogen becomes an excellent conductor, like a metal.  The main uncertainty concerning the hydrogen EOS is in the region of 0.5--10 Mbar (50 -- $10^3$ GPa), where the  transition from the molecular phase to the metallic phase occurs.  
 In fact, the metallization of hydrogen is an active area of research and the exact metallization pressure/temperature is still being debated, but is expected to be at $\sim$1 Mbar for Jupiter's conditions\footnote{It should also be noted 
 that the nature of the transition of hydrogen from "molecular" to "metallic" along Jupiter's and Saturn's adiabats is still debated. 
 The transition could be a "first order" or smooth, although most studies, imply that in Jupiter's and Saturn interiors the transition is smooth and is a first order transition at  lower (intermediate) temperatures.}.  
Finally, it is clear the Saturn's adiabat covers lower temperatures and pressures (due to its lower mass). As a result, Saturn's interior consists of a smaller fraction of metallic hydrogen in comparison to Jupiter, and since this regime of the EOS is less understood there is less uncertainty in Saturn's structure due to the hydrogen EOS (Saumon \& Guillot, 2004). On the other hand, as we discuss below, Saturn is more likely  to be affected by the phase separation of helium. 

\begin{figure}[h!]
\centering
{\includegraphics[scale=.4]{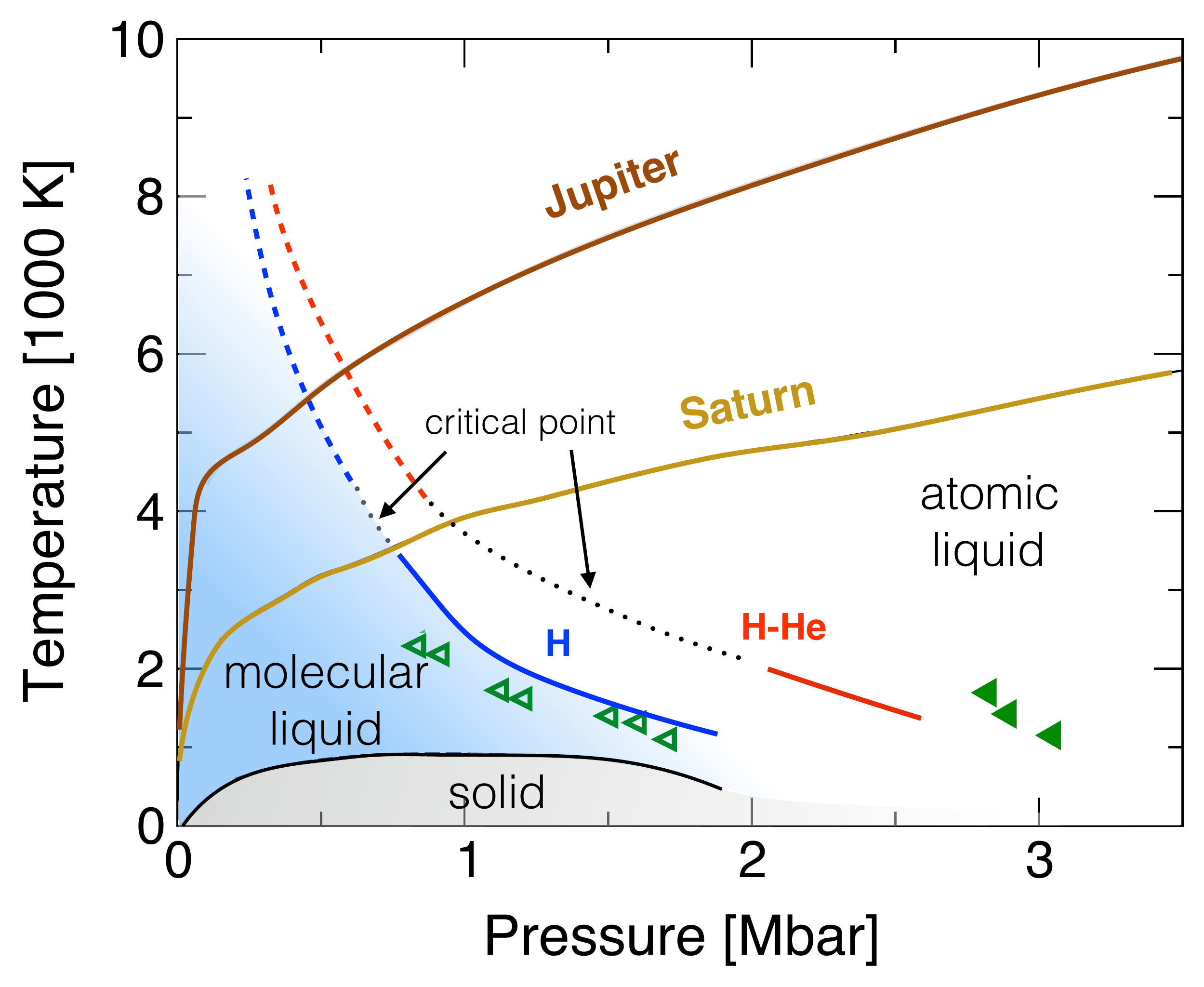}}
\vskip -9pt
{ \caption{\small Phase diagram of hydrogen. 
Grey area: solid phase (experimental),
shaded blue: insulating molecular liquid.
solid blue: first-order liquid-liquid phase boundary (QMC)
dashed blue: continuos liquid-liquid phase boundary (QMC)
black points: possible location of the critical point (end-point of the first-order line)
solid red: first-order for the H-He (QMC)
dashed red: continues liquid-liquid boundary H-He (QMC)
empty triangles: static compression experiments H (Zaghoo et al., 2016, Otha et al., 2015)
solid triangles: dynamic compression H (Knudson et al., 2015).
The figure is based on Mazzola et al., 2018.}}
\end{figure}

\subsubsection{Helium and hydrogen}
The behaviour of pure helium at the extreme conditions inside giant planet interiors is more constrained than that of hydrogen simply because helium ionization requires larger pressures and a phase transition is not expected to take place.
In Figure 1, the blue and red curves correspond to the phase diagram of pure hydrogen and a hydrogen-helium mixture (red) with a proto-solar value, respectively (see Mazzola et al.~2018 for details).  
It is clear that the presence of helium delays the dissociation (metallization) pressure, compared to pure-hydrogen and therefore, the presence of helium cannot be neglected when estimating the metallization  pressures in Jupiter/Saturn interiors.  

In addition, the interaction between hydrogen and helium under the interior conditions of Jupiter and Saturn leads to challenges in determining the EOS.  
This is because helium is expected to become immiscible in hydrogen leading to helium settling (known as "helium rain") that results in a non-homogenous distribution of helium within the planet where helium settles (and is therefore enriched) toward the deep interior.  This phenomenon of "helium rain" was predicted already decades ago (e.g., Salpeter, 1973, Stevenson, 1975, Stevenson \& SalPeter 1977a,b), and got observational support when the helium in the atmospheres of Jupiter and Saturn was found to be depleted in comparison to the proto-solar value. Recently, {\it ab initio} calculations of the phase diagram have confirmed the immiscibility of helium in hydrogen (Salpeter, 1973, Lorenzen et al., 2011, Morales et al., 2013, Sch{\"o}ttler \& Redmer, 2018 and references therein). Figure 2 shows the phase diagram for a H-He mixture with a helium mole concentration slightly lower than proto-solar (see Guillot \& Gautier 2014 for details). 
The exact location in the phase diagram in which helium rain occurs is still being investigated, and is of great importance for understanding the structure and evolution of both Jupiter and Saturn.  
What seems to be robust is that since Saturn has a smaller mass than Jupiter, and therefore its internal temperatures and pressures are lower, it is located  "deeper" within the phase diagram in comparison to Jupiter (see Figure 2). This means that the process of helium rain is more significant in Saturn and has begun earlier. Jupiter's interior on the other hand, has to cool for a longer time to reach the temperatures corresponding to this phase separation. This is consistent with the measurements of helium in the giant planet atmospheres where Saturn's atmosphere is found to be more depleted in helium.  

\begin{figure}
\centering
{\includegraphics[scale=.25]{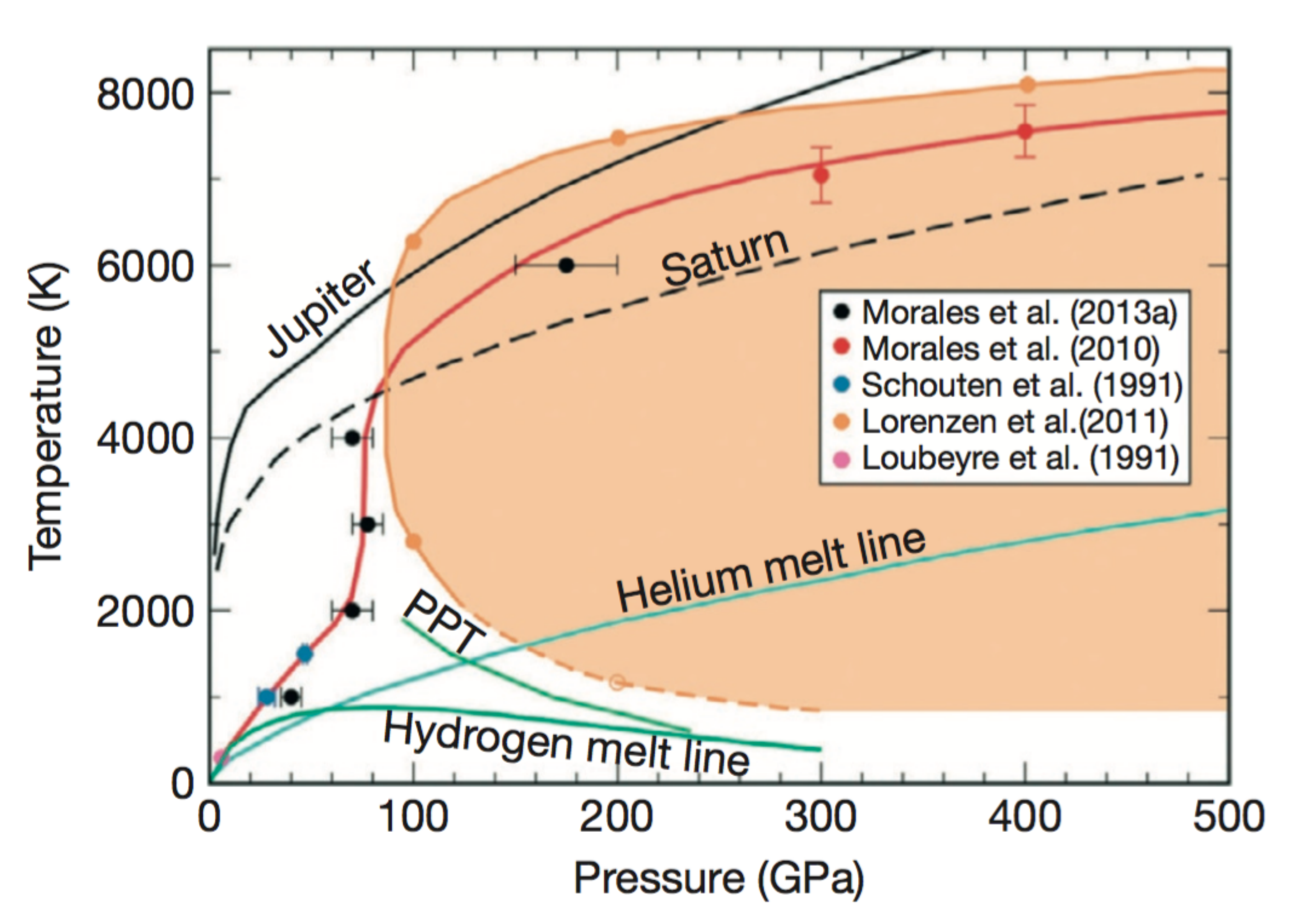}}
\vskip -9pt
{ \caption{\small Phase diagram for a hydrogen-helium mixture. The orange region shows the region of  the H-He separation as derived by Lorenzen et al.~(2011). 
The red curve shows the critical temperature for that separation according to Morales et al. (2013a). Numerical and experimental results by Schouten et al.~(1991) and Loubeyre et al.~(1991) are also presented. 
The back curves show the isentropes of Jupiter (plain) and Saturn (dashed), respectively. The figure is taken from Guillot \& Gautier, 2014.}}
\end{figure}

\newpage
\subsubsection{Heavy elements}
In astrophysics, heavy elements represent all the elements that are heavier than helium. Ideally, structure models should include all the possible elements when modelling the planetary interior. 
However, this introduces an additional complexity to the models, because the ratios between the different elements within the planets have to be assumed. In addition, the details of the EOS of the heavy elements are less crucial, since the temperature dependence on the density of these elements is rather weak at giant planet interior conditions, and their contribution to the planetary density is of a second order effect in comparison to H-He (e.g., Baraffe et al., 2008; Saumon \& Guillot, 2004; Fortney et al., 2018). 
Often, the heavy elements in Jupiter and Saturn are represented by water and/or rock, where rocks lead to about 50\% less massive cores compared to water (e.g., Fortney \& Nettelmann, 2010). While the measured gravitational field is essentially blind to the innermost regions of the planet, these regions can affect the density profile indirectly via the constraints on the outer envelope (e.g., Helled et al., 2011, Guillot \& Gautier, 2014). 
Currently there is ongoing progress in {\it ab-initio} calculations of the EOSs for water, ammonia, silicates and iron as well as their miscibility in metallic hydrogen (e.g., French et al. 2009, Knudson et al., 2012, Wilson \& Militzer, 2010; 2012). 
As such ab-initio EOS calculations become available, it is desirable to include them in structure models and further investigate their effect on the inferred composition and internal structure.  

\section{Internal Structure Models}
In the last few decades many studies aimed to better constrain the interiors of Jupiter and Saturn (e.g., Saumon \& Guillot, 2004; Militzer et al., 2008; 2016; Nettelmann et al., 2008; 2012, 2015; Helled\& Guillot, 2013; Hubbard \& Militzer 2016; Miguel et al., 2016).  
Unfortunately, there is no unique solution for the internal structure of a planet. The non-uniqueness nature of the problem is inherent because the available data are (and will remain) insufficient to uniquely infer the planetary internal structure. 
In addition, the inferred structure depends on the model assumptions and the EOSs used by the modeller.  
The main uncertainties in structure models are linked to the following assumptions/setups: 
(i) the number of layers (ii) the composition and distribution of heavy elements (iii) the heat transport mechanism, (iv) the transition pressure of hydrogen metallisation and (v) the rotation period and the dynamical contribution of winds (e.g., differential rotation). 
\begin{figure} [h!]
\centering
{\includegraphics[scale=.46]{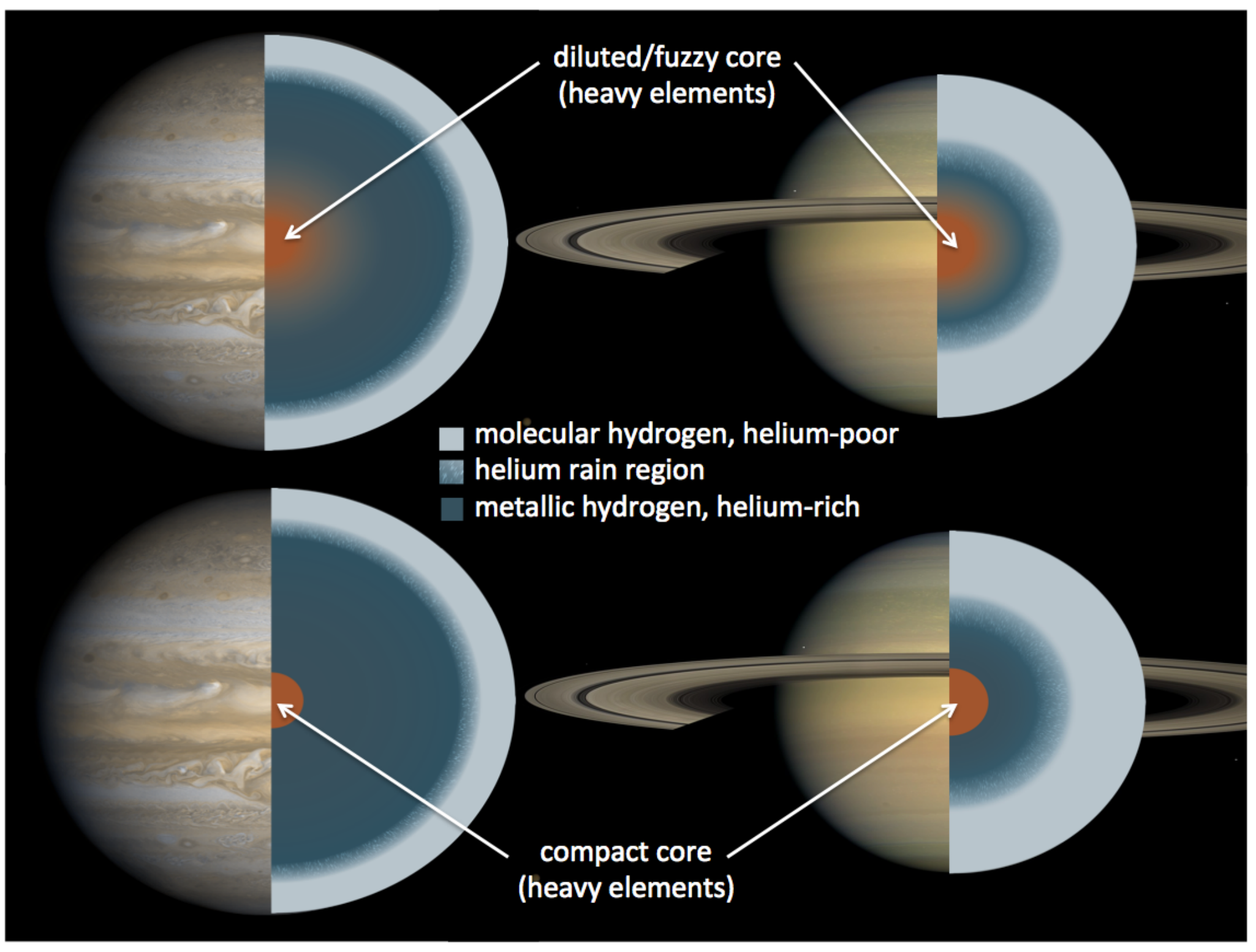}}
\vskip -9pt
{ \caption{\small Sketches of the internal structures of Jupiter and Saturn as inferred from structure models. For each planet two possible structures are shown: one consisting of distinct layers and one with a 
gradual distribution of heavy elements.  Schematic representation of the interiors of Jupiter and Saturn. 
The core masses of Jupiter and Saturn are not well-constrained, for Saturn, the inhomogeneous region could extend down all the way to the center resulting in a "helium core".}}
\end{figure}

Typically, the interiors of Jupiter and Saturn are  modelled assuming the existence of a distinct heavy-element core which is surrounded by an inner envelope of metallic hydrogen and an outer envelope of molecular hydrogen. 
Due to the indication of helium rain in the planets, the inner and outer envelopes are set to be helium-rich and helium-poor, respectively. 
For the heavy elements distribution there are two common assumptions. In the first, they are assumed to be homogeneously mixed within the two envelopes. 
Then, if $Z_{in}$ and $Z_{out}$ represent the heavy element mass fraction in the inner and outer envelopes, respectively, for this case $Z_{in} = Z_{out}$.  
In the second case, the heavy element enrichment is assumed to be higher at the metallic region (inner envelope), i.e., $Z_{in} > Z_{out}$.  In both of these cases the heavy elements are taken to be homogeneously distributed, suggesting a homogenous composition, at least within one part of the envelope. 
The location where the envelope is divided to a helium-poor/helium-rich region corresponds to the pressure in which helium becomes immiscible in hydrogen.  
For simplicity,  for the models with $Z_{in} \neq Z_{out}$ the location of the heavy element discontinuity is assumed to occur at the same location. 
\par

The division of the planetary structure into three layers is not written in stone, it only represents the simplest model that can be considered. 
It may indeed be that the discontinuity in helium and the heavy elements occurs at different pressures, and especially in the case of Saturn, that a nearly pure-helium layer also exists (e.g., Fortney \& Hubbard, 2003). 
In addition, the core itself might not be a distinct region (as well as non-existing), and the heavy-elements might have a gradual distribution along the planetary interior.   
Even within this simple 3-layer model framework, the inferred composition and core mass depend on the model assumptions. More complex models increase the range of possible solutions even further. 
Figure 3 shows a simple sketch of the interiors of Jupiter and Saturn. Figure 4 shows representative density and pressure profiles within the planets for standard 3-layer models.

\begin{figure}[h!]
\centering
{\includegraphics[scale=.6626]{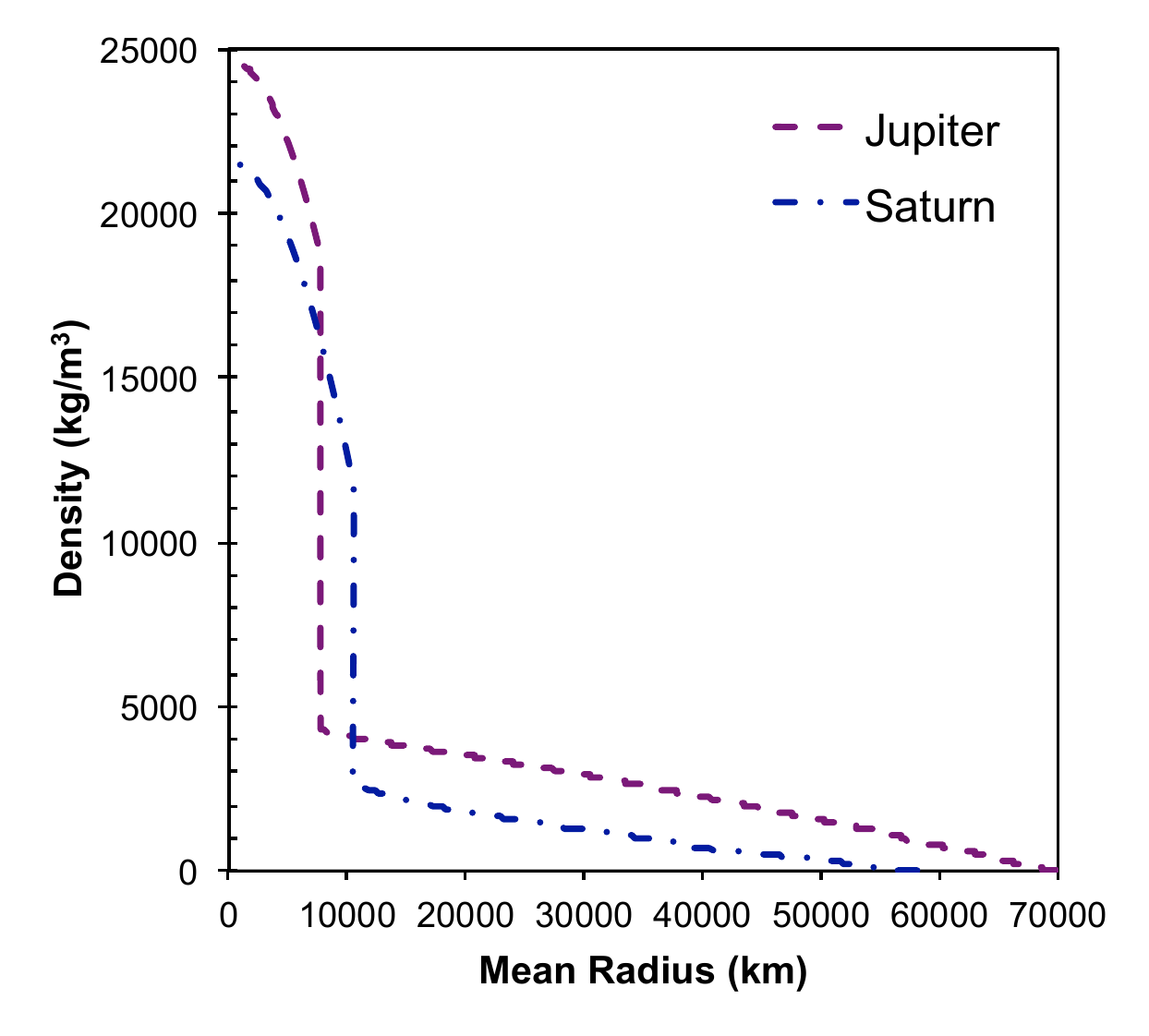}}
{\includegraphics[scale=.6626]{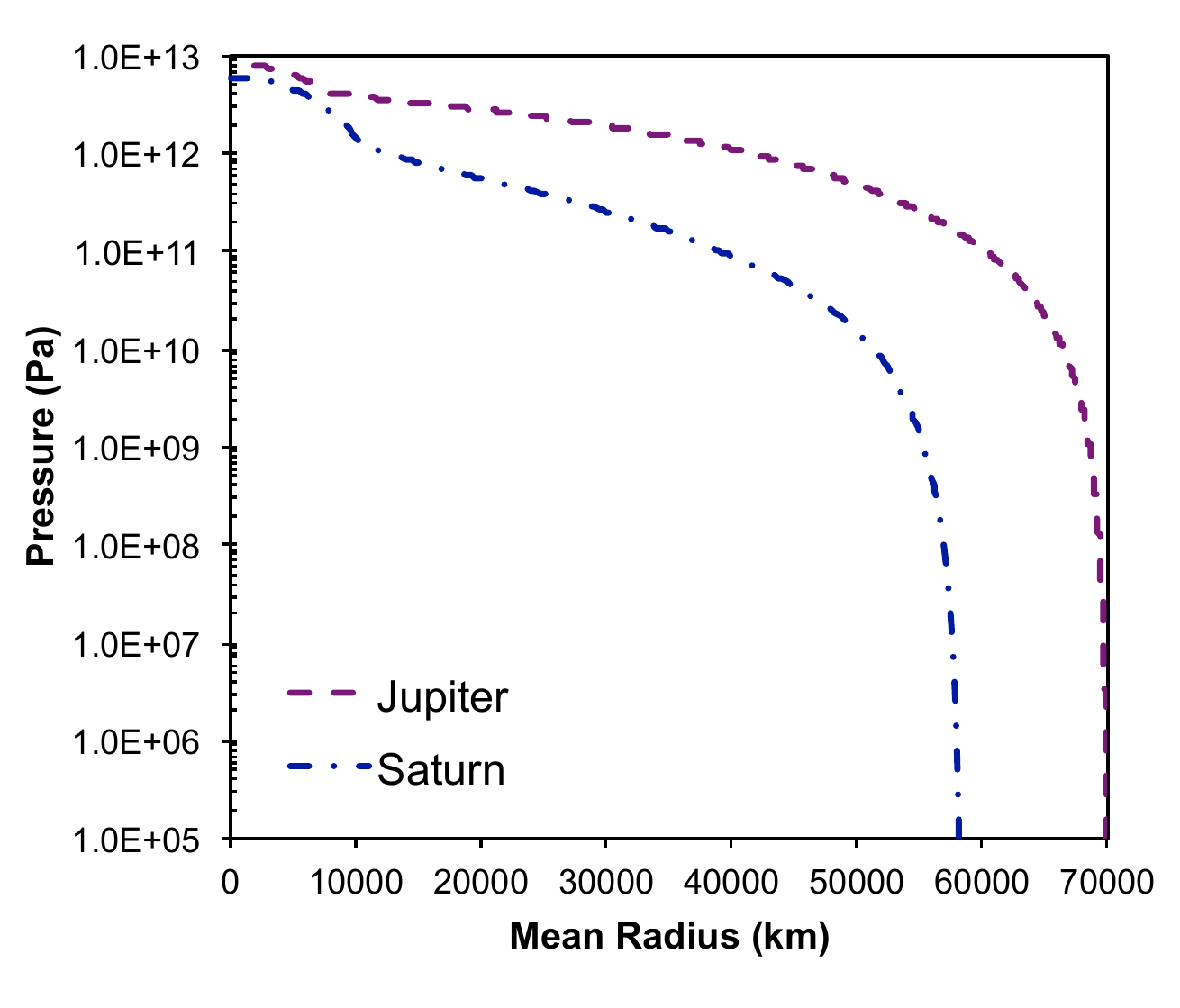}}
\vskip -9pt
{ \caption{\small Representative density (left)  and pressure (right) profiles of Jupiter and Saturn as a function of the planetary mean radius. 
The data are taken from Miguel et al. (2016) and Helled \& Guillot (2013), respectively).}}
\end{figure}

\subsection{Jupiter} 
For Jupiter, structure models typically differ by the assumption regarding the heavy element distribution, the assumed number of layers and the calculated entropy from EOS calculations. 
Saumon \& Guillot (2004) explored the possible range of solutions for Jupiter using different EOSs and inferred total heavy-element mass between 10 and 40 M$_{\oplus}$, and core masses between 0 and 10 M$_{\oplus}$. 
Later, when DFT EOS calculations became available, new Jupiter models have been presented. The first set of models was based on the entropy calculation of the Rostock H-He EOS which was calculated using {\it ab-initio} DFT (e.g., Nettelmann et al.~2008; 2012;   Becker et al.~2014). 
In Nettelmann et al.~(2008), Nettelmann(2012), Becker et al.~(2014), and Nettelmann (2017) the models relied on the 3-layer assumption, where the interior is separated into a distinct core and two homogeneous envelopes. 
The helium mass fraction in the outer envelope, $Y_{out}$, was set to match the Galileo entry probe value of $Y$ = 0.238. The inner envelope helium abundance $Y_{in}$ is chosen to yield a bulk helium mass fraction that reproduces  
the proto-solar value. 
The heavy element mass fractions $Z_{out}$ and $Z_{in}$ were chosen to match the then measured values of the low-order gravitational harmonics $J_2$ and $J_4$, with $J_4$ being slightly more sensitive to $Z_{out}$ than $J_2$ is and vice versa for $Z_{in}$. 
In the transition between the inner and outer envelopes the pressure and temperature are assumed to change continuously, while the density and entropy have discontinuities.  
In these models, the transition pressure $P_{trans}$ is taken to be a free parameter, between 1 and 5 Mbar, although H-He phase diagram calculations suggest that $P_{trans}\sim 1$ Mbar. Higher assumed $P_{trans}$ values lead to higher envelope metallicity and smaller inferred values for the core mass. Overall, the studies of this group have confirmed the ranges derived by Saumon \& Guillot (2004), where the core mass was smaller than $\sim$10 M$_{\oplus}$ with a global enrichment of tens M$_{\oplus}$ of heavy elements. 
\par

A second type of Jupiter models were based on B.~Militzer's H-He EOS calculation, also using {\it ab initio} DFT-MB (e.g., Militzer et al., 2008; Hubbard \& Militzer, 2016). A comprehensive equation of state of H-He mixtures and their inferred internal energies as well as a Jupiter adiabat have been presented by Militzer \& Hubbard (2013). These models lead to a significant inferred core mass for Jupiter, of the order of 15-20 M$_{\oplus}$ with a low envelope metallicity, sometimes even less than solar.  
Note that in these Jupiter models the transition pressure is not taken as a free parameter but is set by identifying the location in which the pressure of Jupiter's adiabat intersects with the H-He immiscibility region as derived by Morales et al., 2013 (see Figure 2).  
It should be noted that the EOSs used by Militzer and collaborators and Nettelmann and collaborators are not very different in terms of the raw data, but in the entropy calculation, and therefore in the constructed adiabat. A hotter adiabat, as inferred by  Nettelmann and collaborators leads to a larger inferred heavy-element mass (see Militzer et al., 2016 and Miguel et al. for further discussion). 
\par

The recent gravity measurements from the Juno spacecraft (Bolton et al., 2017; Iess et al. 2018) introduced new constraints on Jupiter structure models. 
Jupiter models that fit the Juno data have been presented by Wahl et al.~(2017), Nettelmann (2017), and Guillot et al.~(2018).   
Overall, it seems that preferable solutions are ones with cores ($\sim$10 M$_{\oplus}$) and a discontinuity of the heavy-element enrichment in the envelope, with the inner helium-rich envelope consists 
of a more heavy elements than the outer, helium-poor envelope (i.e., $Z_{in} > Z_{out}$). 
In addition, interior models of Jupiter that fit Juno data suggest that another feasible solution for Jupiter's internal structure is the existence of a diluted/fuzzy core (e.g., Wahl et al., 2017).  
In this case, Jupiter's core is no longer viewed as a pure heavy-element central region with a density discontinuity at the core-envelope-boundary, but as a central region that its composition is dominated by heavy elements, which could be gradually distributed or homogeneously mixed. Such a diluted core could extend to a few tens of percents of the planet's total radius, and can also consist of lighter elements (H-He). 
While the total amount of heavy elements in the central region of a diluted core models in the central region does not change much (e.g., Wahl et al., 2017, Nettelmann, 2017), the size of the core increases significantly due to the inclusion of H-He.   
\par

The existence of a diluted core, or a steep heavy-element gradient inside Jupiter is actually consistent with formation models of Jupiter (see section 4.3 for details). 
Giant planet formation models in the core accretion scenario (e.g., Pollack et al., 1996) suggest that once the core mass reaches $\sim 1-2 M_{\oplus}$ the accreted solid material (heavy elements) vaporise and remain in the planetary envelope (e.g., Stevenson, 1982). This leads to a structure in which the deep interior is highly enriched with heavy-elements, with no sharp transition between the core and the inner envelope (e.g., Helled \& Stevenson, 2017 and references therein). Another explanation for a diluted core is core erosion. If the heavy elements within a compact core are miscible in metallic hydrogen (e.g., Wilson \& Militzer, 2010, 2012), the presence of vast convection could mix some of the core elements in the deep interior (e.g., Guillot et al. 2004). As discussed below, long-term evolution models of Jupiter with composition gradients suggest that steep composition gradients can persist up to present-day (see section 3.4). 
\par

Finally, it is important to note that the inferred core mass and total planetary enrichment do not only depend on the assumed EOS but also on the model assumptions. Jupiter's structure models as presented by Wahl et al.~(2017) show that both fuzzy and compact cores are consistent with the Juno data, with the  core mass being between $\sim$ 1.5 and 20 M$_{\oplus}$, depending on the model. A Jupiter structure model with a diluted core resembles the primordial structure derived by formation models (Stevenson 1985; Helled \& Stevenson, 2017), 
providing a potential link between giant planet formation models, and current-state structure of the planets. In all of these models, Jupiter's normalized moment of inertia was found to be $\sim$0.264. This value is now well-constrained, at least from the modelling point of view, due to the accurate determination of Jupiter's gravitational field by Juno (e.g., Wahl et al.~2017). Indeed, it has been shown that there is a very strong correlation between $J_2$ and the moment of inertia but this is not a perfect one-to-one correspondence (e.g., Helled et al., 2011 and references therein).    
\par

\subsection{Saturn}
Saturn is often considered as a small version of Jupiter but in fact the two planets have significant differences. 
First, the relative enrichment in heavy elements is rather different, as well as the geometry the magnetic field, the axis tilt, and the long-term evolution. 
Just from a simple comparison of their normalized moment of inertia values one can conclude that Saturn is more centrally condensed  compared to Jupiter. 
Naively, one would expect that it is easier to model Saturn's interior since a smaller portion of its mass sits in the region of the high uncertainty in the hydrogen EOS, but this is not the case, due to the possibility of helium rain. 
Additional complication arises from the uncertainty in Saturn's rotation period and shape (see Fortney et al., 2017 and references therein).  

Saturn models calculated by Helled \& Guillot (2013) also used the 3-layer model approach, the range of the helium mass fraction in the outer envelope $Y_{out}$ was taken to be between 0.11 and 0.25 
with a global $Y$ = 0.265 -- 0.275 consistent with the protosolar value. Here the EOS of H-He was set to the SCVH which was calculated for a large range of pressures and temperatures (Saumon et al., 1995), and has been broadly used in the astrophysics community.  A range of temperatures at 1 bar were considered (130 -- 145 K) and $P_{trans}$ was allowed to range between 1 and 4 Mbars. For the heavy element distribution, they assumed $Z_{in}=Z_{out}$. These Saturn models also accounted for the uncertainty in Saturn's shape and rotation rate/profile (see section 4.1). Saturn models were constructed for two different assumed rotation periods for both the Voyager and Cassini gravity data.  
For the range of different model assumptions, the derived core mass was found to range between $\sim$ 5 and 20 M$_{\oplus}$, while the heavy-element mass in the envelope was found to be between $\sim$ 0 and 7 M$_{\oplus}$. 
Like for Jupiter, increasing $P_{trans}$ leads to smaller core masses and more enriched envelopes. 
\par

Finally, the Cassini gravity data reduces the inferred core mass by about 5 M$_{\oplus}$. It should be noted, however, based on the recent Juno data for Jupiter as well as new studies on giant planet formation and evolution, that the assumption of $Z_{in}=Z_{out}$ might be inappropriate, and a more realistic assumption is $Z_{in} > Z_{out}$.
%
The total heavy-element mass in Saturn is estimated to be $\sim$16 - 30 M$_{\oplus}$ with a core mass between zero and 20 M$_{\oplus}$ (e.g., Saumon \& Guillot, 2004; Nettelmann et al., 2012; Helled \& Guillot, 2013). 
However, as discussed above, this conclusion is based on relatively simple interiors models. 
Figure 5 provides a schematic presentation of the two possible internal structures of Jupiter and Saturn. 

\begin{figure}
\centering
{\includegraphics[scale=.565]{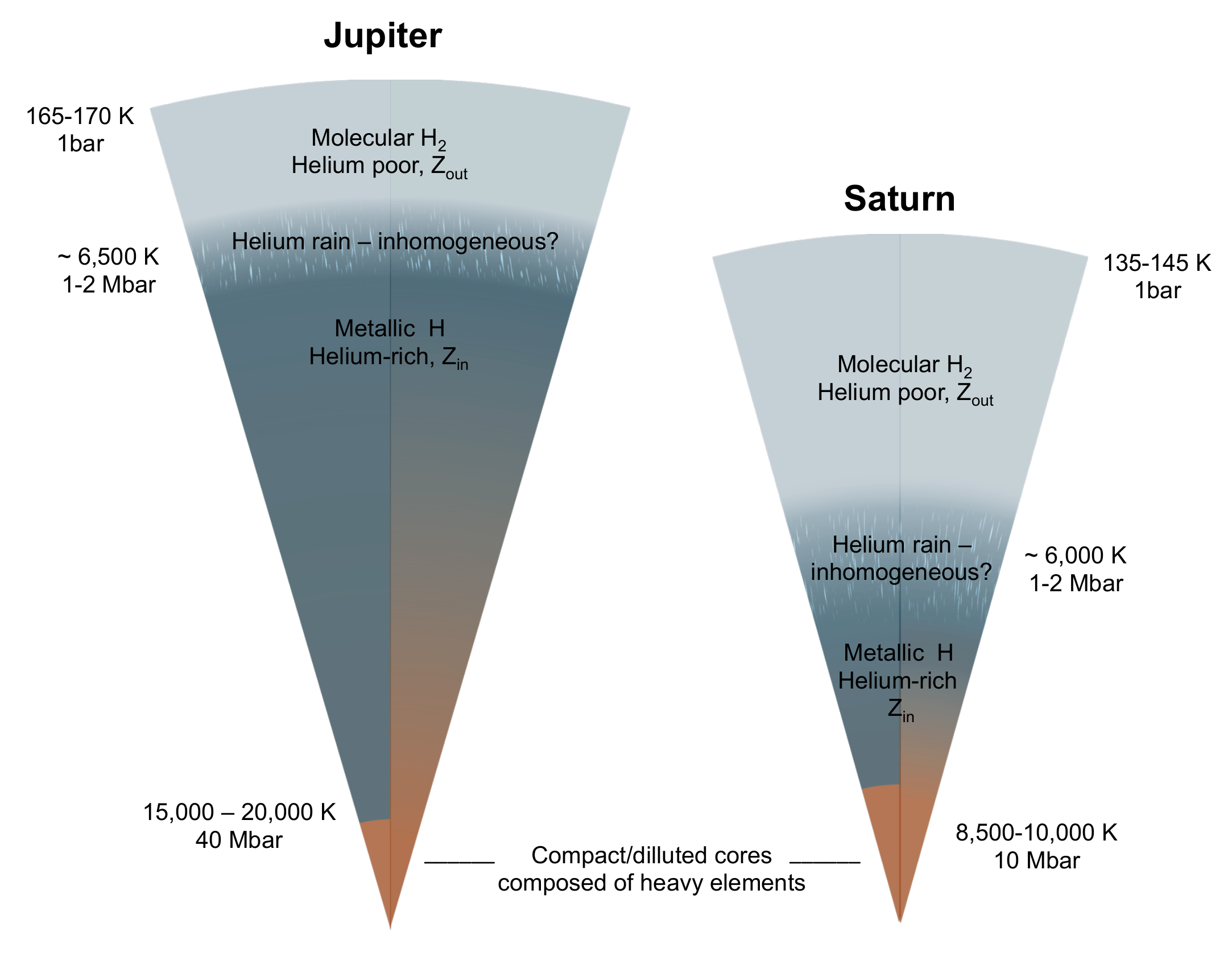}}
{ \caption{\small Sketches of the internal structures of Jupiter and Saturn. }}
\end{figure}
\newpage
\subsection{Non-Adiabatic Interiors}
Standard structure models of Jupiter and Saturn assume that the dominating energy transport mechanism is convection, i.e., that the temperature gradient is given by the adiabatic one, apart from the (thin)  outer radiative atmosphere. 
This assumption simplifies the calculation since the temperature profile is then well-constrained, and in addition, one can assume that the composition within the envelope(s) is homogenous. 
However, we now realise that in some cases (and perhaps in most cases) a fully adiabatic model for the giant planets is too simplistic. 
Non-adiabatic giant planet interiors are in fact a natural outcome of their formation process where the accreted heavy elements result in a non-homogenous interior (e.g., Stevenson 1985, Helled \& Stevenson, 2017, Lozovsky et al., 2017). 
Non-adiabatic interiors can also be a result of core erosion (e.g., Guillot et al., 2004) and immiscibility of materials in metallic hydrogen (e.g., Wilson \& Militzer, 2011; 2012; Soubrian \& Militzer, 2016). 
\par

The existence of composition gradients can inhibit convection due to their stabilising effect. Moderate composition gradients can be erased by overturning convection, especially at early evolution stages where convection is strong, which leads to a rapid mixing and homogenization of the planet. 
Otherwise they can either lead to layered-convection, a less efficient type of convection (e.g., Wood et al., 2013), or inhibit convection and lead to heat transport by conduction/radiation.
\par

Leconte \& Chabrier (2012, 2013) accounted for the possibility of double-diffusive convection in both Jupiter and Saturn interiors caused by heavy-element gradients. 
It was shown that both Jupiter and Saturn can satisfy all the observational constraints also when assuming non-adiabatic structures with compositional gradients throughout the entire planetary interiors. 
Since in this scenario heat loss (cooling) is less efficient, the planetary interiors can be much hotter, and the planets can accommodate larger amounts of heavy elements. 
The core masses  derived by these models were found to be 0--0.5 M$_{\oplus}$ for Jupiter and $\sim$ 10--21 M$_{\oplus}$ for Saturn. The heavy element mass in the envelope was found to be 
and 41--63.5 M$_{\oplus}$ and  10-36 M$_{\oplus}$ for Jupiter and Saturn, respectively (see Leconte \& Chabrier, 2012 for further details). Although these models could be viewed as extreme cases since the composition gradients are assumed to persist  across the entire planetary interiors, they clearly demonstrate the importance of the model assumptions and the limitation of the simple 3-layer models. 
It is also interesting to note that although the semi-convective structure models for Jupiter and Saturn are much more metal-rich, the solution for Jupiter indicates the absence of a core.   
\par 

Evidence of a non-adiabatic interior for Saturn is also indicated from the observed frequency spectrum of its ring oscillations. 
Some of Saturn's ring modes observed by the Cassini spacecraft can be attributed to oscillations within the planetary interior 
(Hedman \& Nicholson, 2013). 
An analysis of the splitting of these oscillation modes suggests the existence of a thick stably-stratified region above the core where gravity modes can penetrate (Fuller, 2014). 
Currently, this is the only proposed explanation to the unexpected splittings, via interactions between f-modes propagating in the convective envelope and g-modes propagating in the stable region of the deep interior.
While further investigations on this topic are required, these important observation and analysis further suggest that a fully convective structure is too simplistic for describing Saturn's (and possibly Jupiter's) interior.

\subsection{Evolution Models}
Another piece of information that can be used to constrain structure models is the planetary evolution. 
The idea is that the current-state structure of the planets must be consistent with the age of the Solar-System ($\sim$ 4.56$\times10^9$ years), i.e., with the planetary evolution.  
In fact, the simple assumption of an adiabatic structure was originated by an evolution model where it was shown that the high thermal emission of Jupiter is somewhat consistent with an interior that is convective (e.g., Hubbard, 1968, Guillot et al., 1994;   Fortney et al. 2011).  
Evolution models with layered convection in the helium-rain region of Jupiter and Saturn have recently been calculated (Nettelmann et al. 2015, Mankovich et al. 2016). In these models, the molecular envelope cools over time, but the deep interior can even heat up if the superadiabaticity in the inhomogeneous He-rain zone is strong. While it is not yet clear whether layer-convection occurs in the helium mixing region\footnote{This would depend on the thermodynamic behavior of the hydrogen-helium mixture in the presence of a phase separation.}, these models show that non-adiabaticity is an important aspect that should be considered when calculating the long-term evolution of gaseous planets such as Jupiter and Saturn.   

Evolution models with primordial composition gradients for Jupiter and Saturn have also been presented (Vazan et al., 2016, 2018). It was found  
that a moderate primordial heavy-element gradient becomes homogenous via convective mixing after several million years and that this mixing leads to an enrichment of the planetary envelope with heavy elements. 
On the other hand, if the primordial composition gradient is steep  convection in the deep interior was found to be inhibited. 
This affects the thermal evolution, and leads to hotter interiors in comparison to the standard adiabatic case. Like in the structure models with layered convection, also here the total heavy element mass in 
the planets is higher than in the adiabatic models, and was found to be up to 40 M$_{\oplus}$ and 36 M$_{\oplus}$ for Jupiter and Saturn, respectively. 

\section{New Insights from the Juno and Cassini missions}
In July 2016 the Juno mission began to orbit Jupiter and among other things, has provided an accurate measurement of Jupiter's gravitational and magnetic fields (e.g., Bolton et al., 2017, Folkner et al., 2017, Iess et al., 2018).  
At the same time, the Cassini spacecraft performed its last orbits having geometries similar to that of Juno, known as  Cassini Grand Finale, providing similar information about  Saturn's fields (e.g., Spilker 2012),    
allowing a comparative study of the Solar-System giant planets.  
Studies and investigations are still ongoing, and more results are likely to appear at the time of (or after) this review but some key conclusions and new insights on the structure of Jupiter and Saturn from these recent measurements have already been achieved. 

\subsection{Rotation Rate and Depth of Winds}

The atmospheres of both Jupiter and Saturn have strong zonal winds with equatorial speeds of $\sim$ 100 m s$^{-1}$ and 400 m s$^{-1}$, respectively.  
These zonal wind velocities are relative to the assumed rotation period of the planet's deep interior (Table 1). In fact, it is not necessarily intuitive to think that giant planets rotate as solid bodies, 
and are therefore represented by a single rotation period, due to the fact that they are fluid objects and are characterised by zonal winds that hint the possibility of differential rotation (on cylinders). 

Jupiter's rotation period is assumed to be represented by the rotation period of its magnetic field which is tilted $\sim$10 degrees from its spin pole, and has not changed in many decades (e.g., Riddle and Warwick, 1976; Higgins et al., 1996). 
On the other hand, Saturn's magnetic pole is aligned with its rotation axis.  
This spin-aligned configuration and the fact that the magnetic field is dipolar prevent a direct determination of the rotation rate of Saturn's deep interior because there is no variable component of the magnetic field that is associated with   the planetary rotation (e.g., Cao et al. 2011; 2012). 
The 10h 39m 22s rotation period of Saturn, which leads to an equatorial speed of  400 m s$^{-1}$ was derived  from the Voyager spacecraft measurement of  the periodicity in Saturn's kilometric radiation (e.g., Ingersoll \& Pollard, 1982; Dessler, 1983).  
In fact, measurements from the Cassini spacecraft did not only measure a different periodicity by several minutes but also showed that the period is changing with time (e.g., Gurnett et al., 2007), suggesting that the periodicity in Saturn's kilometric radiation does not 
represent the rotation of the deep interior.  Therefore, at the moment Saturn's rotation period is not well-constrained.  
Accordingly, the atmospheric zonal wind velocities with respect to the underlying rotating planet are also unknown for Saturn. 
Several theoretical approaches have been presented to constrain Saturn's rotation period, and the estimated values range between $\sim$ 10 hr 32m and 10h 45m (e.g., Anderson \& Schubert 2007; Read et al., 2009; Helled et al., 2015; Mankovich et al. 2018). 
While an uncertainty of about ten minutes sounds small, it can affect the inferred internal structure of the planet, and also has implications on its atmosphere dynamics. 
\par 

The relation between the rotation period of Jupiter and Saturn to their zonal wind, physical shapes, and gravitational and magnetic fields has been 
studied for decades, and is still being investigated. Nevertheless, recently substantial progress in this direction has been made thanks to the Cassini and Juno missions. 
Deep winds can change the planetary density profile and therefore contribute to the measured gravitational harmonics, and as a result this contribution has to be accounted for 
 as an uncertainty in structure models, since they are hydrostatic, and do not include dynamical effects. 
 The depth of the winds can be constrained by accurate measurements of the high-order gravitational and/or the odd harmonics (e.g., Hubbard, 1999; Kaspi et al., 2010; 2018).  
A determination of the depth of the winds in Jupiter was recently possible thanks to the Juno data (Iess et al., 2018, Kaspi et al., 2018; Guillot et al. 2018). The winds were found to penetrate to  
depths of 2000--3000 km, suggesting that 1\% of the outer planetary mass rotates differentially in patterns similar to that of the observed atmospheric winds. This depth is consistent with the one expected from ohmic dissipation constraints linked to the metallisation of hydrogen (e.g., Liu et al., 2008). 
Since the metallisation in Saturn occurs at deeper regions (due to its smaller mass and resulting pressures), then by following the same argument the depth the winds in Saturn is predicted to reach deeper, down to $\sim$ 9000 km. 
These estimates correspond to depths of around 95\% and 80\% of the total planetary radius for Jupiter and Saturn, respectively (e.g., Cao \& Stevenson, 2017). 

\subsection{Magnetic Fields}
Both Jupiter and Saturn possess dipolar intrinsic magnetic fields. The existence and nature of the magnetic fields provide important observational constraints on their present-day interior structure and dynamics. 
The existence of intrinsic magnetic field requires large-scale motions in a medium that is electrically conducting (e.g., Robert \& King, 2013). 
For Jupiter and Saturn, large-scale radial motions are caused  by the convective motions which also transport heat from the deep interior towards the outer regions with the 
conducting material being metallic hydrogen. 
Indeed, a significant electrical conductivity is expected inside Jupiter and Saturn before the full metallization of hydrogen at Mbar (100 GPa) pressures (e.g., French et al., 2012). The magnitude of the electrical conductivity inside Jupiter and Saturn  combined with the planetary measured magnetic field strength and surface luminosity, can be used to estimate the internal Ohmic dissipation and introduce additional constraints for structure models (e.g., Liu et al., 2008; Cao \& Stevenson, 2017).  
Thus, our ability to decode the interiors of Jupiter and Saturn from the measured properties of the magnetic field is limited by our current understanding of the dynamo process.   
\par

Jupiter's intrinsic magnetic field is strongest among all Solar System planets, with surface field strength ranging from 4 Gauss to 20 Gauss (Connerney et al., 2018; Moore et al., 2018). Recent Juno observations revealed several surprising factors in the morphology of Jupiter's magnetic field. 
When viewed at the Òdynamo surfaceÓ, Jupiter's magnetic field is characterised by an intense isolated magnetic spot near the equator with negative flux, an intense and relatively narrow band of positive flux near 45 degrees latitude in the northern hemisphere, and relatively smooth magnetic field in the southern hemisphere. The north-south dichotomy in Jupiter's magnetic field morphology has been speculated to be due to the existence of a diluted core inside Jupiter, which either limits the dynamo action to the upper layer of Jupiter or creates two spatially separate dynamo action inside Jupiter (Moore et al., 2018). This provides a nice link between internal structure models that are based solely on the gravity data and magnetic field measurements. 
\par

Saturn's intrinsic magnetic field is unusually weak, with surface field strength ranging from 0.2 Gauss to 0.5 Gauss (Dougherty et al., 2005; Cao et al., 2011; 2012). 
Surprisingly, Saturn's magnetic field seems to be perfectly symmetric with respect to the spin-axis (Cao et al. 2011; 2012). Both the weak strength and the extreme spin axisymmetry of Saturn's magnetic field was attributed to the helium rain, which could create a stably stratified layer atop the deep dynamo. However, whether helium rain or composition gradients inside Saturn create substantial stable stratification, 
and whether this stratification layer is above rather than below the deep dynamo is still being investigated. 

\subsection{Constraints on Internal Structure and Origin}
In the standard view of giant planet formation, known as core accretion (e.g., Pollack et al., 1996), a giant planet forms in three stages:
\begin{itemize}
     \item Phase-1: {\it Primary core/heavy-element accretion}. During this early phase, the core accretes solids (planetesimals/pebbles) until it empties its gravitational dominating region (feeding zone). The mass associated with the end of this stage is known as {\it isolation mass} and its exact value depends on the local formation conditions. At this point, the planet is primarily composed of heavy-elements with a negligible fraction of an H-He envelope. 
     \item  Phase-2: {\it Slow envelope accretion}. During this phase, the solid accretion rate decreases, and the H-He accretion rate increases until the envelope accretion rate exceeds the heavy-element accretion rate. 
     The growth of the envelope enlarges the planet's feeding zone and thus allows heavy elements to be accreted but at a slow rate.
     \item   Phase-3: {\it Rapid gas accretion}. Once the H-He mass is comparable to the heavy element mass, the gas accretion rate continuously increases and exceeds the heavy-element accretion rate until the disk can no longer supply gas fast enough to maintain equilibrium and keep up with the planetary contraction and a rapid hydrodynamic accretion of H-He initiates. 
\end{itemize}

In the early core accretion simulations for the sake of numerical simplicity, it was assumed that all the heavy elements reach the core while the envelope is composed of H-He.
However, formation models that follow the heavy-element distribution during the planetary formation show that once the core mass reaches a small value of $\sim$ 1--2 M$_{\oplus}$ and is surrounded by a small envelope, the solids composed of heavy-elements tend to dissolve in the envelope instead of reaching the core (e.g., Lozovsky et al., 2017; Helled \& Stevenson, 2017). In this case, the resulting giant planet has a small core mass and an inner envelope which is enriched with heavy elements. 
Interestingly, this view is consistent with the possibility of Jupiter having a diluted/fuzzy core. Although the prediction on the distribution of heavy-elements corresponds to Jupiter right after its formation, evolution models confirm that in several cases such a structure can persist until today (e.g., Vazan et al., 2016; 2018). A point that still needs to be investigated is whether the composition discontinuity in the heavy elements is caused by the formation process or a result of phase separations and/or core erosion that occur at later stages during the planetary long-term evolution. 
\par

Another missing piece of the puzzle in our understanding of Jupiter is linked to the water abundance. The low abundance of water in Jupiter's atmosphere measured by the Galileo probe is likely to be a result of the special entry spot, which keeps the enrichment of water in Jupiter's atmosphere unknown. 
Jupiter's water abundance is now being measured by Juno using the microwave radiometer (MWR) which probes down to pressure levels of $\sim$ 100 bar at radio wavelengths ranging from 1.3 to 50 cm using six separate radiometers to measure the thermal emissions. 
The water measurement is not only important for constraining Jupiter's origin (e.g., Helled \& Lunine, 2014 and references therein) but also to further constrain structure models. First, since updated Jupiter interior models suggest that  $Z_{in} > Z_{out}$, the water measurement will provide a lower bound for the total (water) enrichment within Jupiter. Second, since several of the new interior models infer a low metallicity for the outer envelope, they could be excluded. 
Finally, the variation of water with depth can teach us about Jupiter's  atmosphere dynamics and put constraints on the convective behaviour in the upper atmosphere and could indicate the presence of a non-convective region within Jupiter's interior. 
It should be kept in mind that the MWR measurement still reveals the information about a very small fraction of the planet, nevertheless, when combined with other measurements, it will provide new insights about the most massive planet in the Solar System. 
Future studies should explore the relations between various formation and evolution model assumptions and the inferred planetary composition and internal structure. 

\section{Summary \& Future Outlook}
There are still many unsolved questions regarding the origin and internal structure of Jupiter and Saturn. As open questions are being solved, new questions arise and our understanding is still incomplete. 
Nevertheless, we are now in a golden era for giant planet exploration given the ongoing Juno mission and the recent measurements from Cassini Grand Finale that are still being processed. The possibility of having similar information of Jupiter and Saturn simultaneously opens the opportunity to improve our understanding of giant planets and to explore the physical and chemical processes that lead to the differences. 
We now know that even within our planetary system, there are significant differences between the two giant planets, suggesting that there is no one simple way to model giant planet interiors. 
 
The continuous theoretical efforts and the new measurements from Juno and Cassini provide data that will keep planetary modellers busy for a while. In the meantime, we also need to keep improving the knowledge of the EOS of the different elements and their interaction, and combine all the available information (gravity field, magnetic field, atmospheric composition, etc.) to further constrain the planetary interior. In addition, it is desirable to develop a united theoretical framework for giant planet formation, evolution and current-state structure. 
\par

Future missions will also play an important role in better constraining the interiors of Jupiter and Saturn. 
The upcoming JUICE mission can reveal further information on Jupiter, and a potential Saturn probe mission will provide constraints on Saturn's atmospheric composition and the process of helium rain. 
Finally, the detection and characterisation of giant planets around other stars, combined with the knowledge of the Solar System giants, can lead to a more comprehensive understanding of gaseous planets. 

\subsubsection*{Acknowledgements}
I thank Nadine Nettelmann, Hao Cao, and Guglielmo Mazzola for their important contributions. 
I also acknowledge valuable comments and support from David Stevenson, Tristan Guillot and Allona Vazan as well as the two anonymous referees. 
Finally, I  acknowledge all the Juno science-team members for inspiring discussions.   

\newpage
\section{References}
{\small
\begin{itemize}
     \item[] Atreya S. K., Mahaffy P. R., Niemann H. B., Wong M. H., Owen T. C.~(2003).  {\it Planet. Space Sci.}, 51, 105\\
    \sitem[] Anderson, J. D.~\& Schubert, G.~(2007). Saturn's Gravitational Field, Internal Rotation, and Interior Structure. {\it Science}, 317, 1384. \\
     \sitem[] Azadi, S.~\& Foulkes, W.~M.~C.~(2013). Fate of density functional theory in the study of high-pressure solid hydrogen. {\it Phys.~Rev.~B}, 88, 014115.\\
     \sitem[] Bahcall, J. N., Pinsonneault, M. H.  \& Wasserburg G. J., (1995). Solar models with helium and heavy-element diffusion. {\it Reviews of Modern Physics} 67, 781Ð808.\\
     \sitem[] Baraffe, I., Chabrier, G., Fortney, J.~\& Sotin, C.~(2014). Planetary Internal Structures. Protostars and Planets VI, Henrik Beuther, Ralf S. Klessen, Cornelis P. Dullemond, and Thomas Henning (eds.), University of Arizona Press, Tucson, 914 pp., 763.\\
     \sitem[] Bolton et al.~(2017). Jupiter's interior and deep atmosphere: the first close polar pass with the Juno spacecraft. {\it Science}, 356, 821--825. \\ 
     \sitem[] Cao, H., Russell, C. T., Christensen, U. R.,  Dougherty, M. K. \& Burton, M. E.~(2011). Saturn's very axisymmetric magnetic field: No detectable secular variation or tilt. {\it Earth and Planetary Science Letters}, 304, 22--28. \\
     \sitem[] Cao, H., Russell, C. T. , Wicht, J., Christensen, U. R.  \& Dougherty, M. K.~(2012). Saturn's high degree magnetic moments: Evidence for a unique planetary dynamo. {\it Icarus}, 221, 388--394.
 \\
     \sitem[] Cao, H.~\& Stevenson, D.~(2017). Zonal flow magnetic field interaction in the semi-conducting region of giant planets, {\it Icarus}, 296, 59--72.\\
     \sitem[] Connerney J.~E.~P.~et al., (2018). A New Model of Jupiter's Magnetic Field from Juno's First Nine Orbits. {\it GRL}, 45, 6. \\
          \sitem[] Conrath, D. \& Gautier, D.~(2000). Saturn Helium Abundance: A Reanalysis of Voyager Measurements. {\it Icarus}, 144, 124. \\
     \sitem[] Debras, F.~\& Chabrier, G.~(2018). A complete study of the precision of the concentric MacLaurin spheroid method to calculate Jupiter's gravitational moments. {\it A\&A},  609, id.A97.\\
     \sitem[] Dessler, A.J.~(1983). Physics of the Jovian Magnetosphere. Cambridge University Press, New York, p. 498.\\
     \sitem[] Dougherty, M.~K., Achilleos, N. Andre, N.~et al.~(2005). Cassini Magnetometer Observations During Saturn Orbit Insertion. {\it Science}, 307, 1266--1270.\\
     \sitem[] Folkner, W.~M.~et al., 2017. Jupiter gravity field estimated from the first two Juno orbits. {\it GRL}, under review. \\
     \sitem[] Fortney, J. J. \& Hubbard, W.~B.~(2003). Phase separation in giant planets: inhomogeneous evolution of Saturn. {\it Icarus}, 164, 228.\\
     \sitem[] Fortney, J.~J.~\& Nettelmann, N.~(2010). The interior structure, composition, and evolution of giant planets. {\it Space Sci.~Rev.}~152, 423. \\
     \sitem[] Fortney, J. J., Ikoma, M.,  Nettelmann, N.,  Guillot, T., and Marley, M. S.~(2011). Self-consistent Model Atmospheres and the Cooling of the Solar SystemÕs Giant Planets. {\it ApJ},  729, 32.\\
     \sitem[] Fortney, J.~J., Helled, R., Nettelmann, N., Stevenson, D.~J., Marley, M.~S., Hubbard, W.~B.~\& Iess, L.~(2016). Invited review for the forthcoming volume "Saturn in the 21st Century." eprint arXiv:1609.06324 \\
     \sitem[] Foulkes, W.~M.~C.,  Mitas, L., Needs, R.~J.~\& Rajagopal, G.~(2011). Quantum Monte Carlo simulations of solids. {\it Rev.~Mod.~Phys.}, 73, 33.\\
     \sitem[] French, M., Becker, A., Lorenzen, W., Nettelmann, N., Bethkenhagen, M., Wicht, J.~\& Redmer, R.~(2012). Ab Initio Simulations for Material Properties along the Jupiter Adiabat.  {\it ApJS}, 202, article id. 5, 11 pp. \\
     \sitem[] Fuller, J.~(2014). Saturn ring seismology: Evidence for stable stratification in the deepinterior of Saturn, {\it Icarus}, 242, 283.\\
     \sitem[] Guillot, T., D. Gautier, G. Chabrier, and B. Mosser, (1994). Are the giant planets fully convective? {\it Icarus}, 112, 337-- 353.\\
     \sitem[] Guillot, T., Stevenson, D.~J., Hubbard, W.~B.~\& D. Saumon (2004). The interior of Jupiter. Jupiter: The Planet, Satellites and Magnetosphere (Cambridge Planetary Science) 
by Fran Bagenal, Timothy E. Dowling \& William B. McKinnon .\\ 
     \sitem[] Guillot, T.~\& Gautier, D.~(2014).  Treatise on Geophysics (Eds. T. Spohn, G. Schubert). Treatise on Geophysics, 2nd edition.  \\
     \sitem[] Guillot, T, . Miguel, Y., Militzer, B.~et al.~(2018). A suppression of differential rotation in JupiterÕs deep interior. {\it Nature}, 555, 227.\\
      \sitem[] Gurnett, D. A., A. M. Persoon, W. S. Kurth, et al.~(2007). The Variable Rotation Period of the Inner Region of Saturn's Plasma Disk. {\it Science} 316, 442.\\
     \sitem[] Hedman, M. M.~\& Nicholson, P. D.~(2013). Kronoseismology: Using Density Waves in Saturn's C Ring to Probe the Planet's Interior. {\it Astronom.~J}, 146, 12.\\
     \sitem[] Helled, R.~\& Stevenson, D., (2017). The Fuzziness of Giant PlanetsÕ Cores. {\it ApJL}, 840, 4. \\
     \sitem[] Helled, R., Galanti, E.~\& Kaspi, Y.~(2015). Saturn's fast spin determined from its gravitational field and oblateness. {\it Nature}, 520, 202--204.\\
     \sitem[] Helled, R.~\& Lunine, J.~(2014). Measuring jupiter's water abundance by juno: the link between interior and formation models. {\it MNRAS},  441, 2273.\\
  \sitem[] Helled, R., Anderson, J.~D., Podolak, M.~\& Schubert G.~(2011). Interior models of Uranus and Neptune. {\it ApJ}, 726,15. \\
     \sitem[]  Helled, R., Bodenheimer, P., Podolak, M. et al.~(2014). Giant Planet Formation, Evolution, and Internal Structure. {\it Protostars and Planets VI}, University of Arizona Press, eds. H. Beuther, R. Klessen, C. Dullemond, Th. Henning. University of Arizona Space Science Series, 914 , 643. \\
     \sitem[] Helled, R.~\& Guillot, T.~(2013). Interior models of Saturn: Including the uncertainties in shape and rotation. {\it ApJ}, 767, 113.\\
       \sitem[] Helled, R., Anderson, J. D., Schubert, G.~\& Stevenson, D.~J.~(2011). Jupiter's Moment of Inertia: A possible Determination by JUNO. {\it Icarus}, 216, 440. \\
     \sitem[] Helled, R.~\& Guillot, T.~(2018).  Internal Structure of Giant and Icy Planets: Importance of Heavy Elements and Mixing. H.J. Deeg, J.A. Belmonte (eds.), Handbook of Exoplanets. \\ 
     \sitem[] Higgins, C.A., Carr, T.D., Reyes, F., 1996. {\it GRL} 23, 2653--2656.\\
     \sitem[] Hubbard, W.~B. \& Militzer, B.~(2016). A preliminary Jupiter model. {\it ApJ}, 820, 80.\\
     \sitem[] Hubbard, W.~B,  Schubert, G., Kong, D. \& Zhang, K.~(2014). On the convergence of the theory of figures. {\it Icarus},  242, 138--141. \\
     \sitem[] Hubbard, W. B., (2013). Concentric Maclaurin spheroid models of rotating liquid planets. {\it ApJ}, 768, 43.\\
     \sitem[] Hubbard, W. B., (2012). High-precision Maclaurin-based models of rotating liquid planets. {\it ApJL}, 756, L15. \\
     \sitem[] Hubbard, W. B.~(1999). Gravitational signature of Jupiter's deep zonal flows. Icarus 137, 357--359.\\
     \sitem[] Hubbard W. B. (1968). Thermal structure of Jupiter. {\it Astrophys. J.}, 152, 745--754.\\
     \sitem[] Iess, L., Folkner, W.~M., Durante, D.~et al., (2018). Measurement of JupiterÕs asymmetric gravity field. {\it Nature}, 555, 220. \\
   \sitem[] Iaroslavitz, E.~\& Podolak, M.~(2007). Atmospheric mass deposition by captured planetesimals. {\it Icarus}, 187, 600.\\
      \sitem[] Ingersoll, A.P., Pollard, D.~(1982). Motion in the interiors and atmospheres of Jupiter
and Saturn--scale analysis, anelastic equations, barotropic stability criterion. {\it Icarus} 52, 62--80.\\
     \sitem[] Knudson, M., Desjarlais, M. P., Becker, et al.~(2015) Direct observation of an abrupt insulator-to-metal transition in dense liquid deuterium. {\it Science}, 348, 1455--1460. \\
     \sitem[] Kaspi,Y., Flierl, G.~R. \& Showman, A.~P.~(2009). The deep wind structure of the giant planets: results from an anelastic general circulation model. {\it Icarus}, 202, 525--542.\\
     \sitem[] Kaspi,Y.,Hubbard,W.B.,Showman,A.P. \&Flierl,G.R.~(2010). Gravitational signature of JupiterÕs internal dynamics.  {\it GRL},  37, L01204.\\
     \sitem[] Kaspi, Y.~(2013). Inferring the depth of the zonal jets on Jupiter and Saturn from odd gravity harmonics. {\it GRL},  40, 676--680.\\
     \sitem[] Kaspi, Y, .Galanti, E., Hubbard, W.~B.~et al.~(2018). Jupiter's atmospheric jet- streams extend thousands of kilometers deep. {\it Nature}, 555, 223.\\
     \sitem[] Leconte, J.~\& Chabrier, G.~(2012). A new vision on giant planet interiors: the impact of double diffusive convection. {\it A\&A}, 540, A20\\
     \sitem[] Leconte, J.~\& Chabrier, G.~(2013). Layered convection as the origin of SaturnÕs luminosity anomaly. {\it Nature Geoscience}, 6, 347.\\
     \sitem[] Liu, J., Goldreich, P. M.~\& Stevenson, D. J.~(2008). Constraints on deep-seated zonal winds inside Jupiter and Saturn. {\it Icarus}, 196, 653--664. \\
     \sitem[] Lorenzen, W., Holst, B.~\& Redmer, R.~(2009). Demixing of Hydrogen and Helium at Megabar Pressures. {\it PRL}, 102(11), 115701.\\
     \sitem[] Lorenzen, W., Holst, B.~\& Redmer, R.~(2011). Metallization in hydrogen-helium mixtures. {\it Phys.~Rev.~B}, 84(23), 235109.\\
     \sitem[] Loubeyre,  P., Letoullec, R.~\& Pinceaux, J.~P.~(1991). A new determination of the binary phase diagram of H$_2$-He mixtures at 296 K. {\it Journal of Physics: Condensed Matter}, 3, 3183.\\
     \sitem[] Lozovsky, M., Helled, R., Rosenberg, E.~D.~\& Bodenheimer, P.~(2017). Jupiter's Formation and Its Primordial Internal Structure. {\it ApJ}, 836, article id. 227, 16 pp.\\
     \sitem[] Mankovich, C., Fortney, J.~J.~\& Moore, K.~L.~(2016). Bayesian Evolution Models for Jupiter with Helium Rain and Double-diffusive Convection. {\it ApJ}, 832, article id. 113, 13 pp. \\
     \sitem[] Mankovich, C.~, Marley, M.~S., Fortney, J.~J.~\& Movshovitz, N.~(2018). Cassini Ring Seismology as a Probe of Saturn's Interior I: Rigid Rotation. {\it ApJ}, submitted. arXiv:1805.10286\\
     \sitem[] Mazzola, G., Helled, R. \& Sorella, S., (2018). Phase diagram of hydrogen and a hydrogen-helium mixture at planetary conditions by Quantum Monte Carlo simulations. {\it PRL}, 120, 025701.\\
     \sitem[] Miguel, Y., Guillot, T.~\& Fayon, L.~(2016). Jupiter internal structure: the effect of different equations of state. {\it A\&A}, 596, id.A114, 12 pp. \\
     \sitem[] Militzer, B., Hubbard, W.~B., Vorberger, J., Tamblyn, I.~\& Bonev, S.~A.~(2008). A massive core in Jupiter predicted from first-principles simulations. {ApJL}, 688, L45.\\
     \sitem[] Militzer, B.~\& Hubbard, W.~B.~(2013). Ab initio equation of state for hydrogen-helium mixtures with recalibration of the giant-planet mass-radius relation, {\it Astrophys. J.}, 774, 148.\\
     \sitem[] Militzer, B., Soubiran, F., Wahl, S.~M., Hubbard, W.~(2016). Understanding Jupiter's interior. {\it JGR: Planets}, 121, 1552-.\\
     \sitem[] Mirouh, G. M., Garaud, P., Stellmach, S., Traxler, A. L., \& Wood, T. S.~(2012). {\it ApJ}, 750, 61.\\
     \sitem[] Moore, K.~M.~Yadav, R.~K.~, Kulowski, L., Cao, H., Bloxham, J., Connerney et al.~(2018). Jupiter's Magnetic Field Morphology: Implications for its Dynamo. {\it Nature}, submitted.\\.
     \sitem[] Morales, M. A., Hamel, S., Caspersen, K.~\&Schwegler,  E.~(2013). Hydrogen-helium demixing from first principles: From diamond anvil cells to planetary interiors. {\it Phys. Rev. B}, 87, 174105.\\
     \sitem[] Morales, M. A.,Schwegler, E., Ceperley, D.~et al.~(2009). Phase separation in hydrogen-helium mixtures at Mbar pressures. {\it PNAS}, 106, 1324.\\
     \sitem[] Nettelmann, N., Fortney, J.~J., Moore, K.~\& Mankovich, C.~(2014). An exploration of double diffusive convection in jupiter as a result of hydrogen-helium phase separation. {\it MNRAS}, 447, 3422.\\
     \sitem[] Nettelmann, N., Helled, R., Fortney, J.~J., and Redmer, R.~(2012). New indication for a dichotomy in the interior structure of uranus and neptune from the application of modi ed shape and rotation data. {\it Planet.~Space Sci.}, special edition, 77, 143.\\
     \sitem[] Nettelmann, N., Holst, B., Kietzmann, A., French, M., Redmer, R., and Blaschke, D.~(2008). Ab initio equation of state data for hydrogen, helium, and water and the internal structure of Jupiter. {\it ApJ}, 683, 1217.\\
     \sitem[] Nettelmann, N., Becker, A., Holst, B.~\& Redmer, R.~(2012). Jupiter models with improved ab initio hydrogen equation of state (H-REOS.2). {\it ApJ}, 750, 52.\\
     \sitem[] Nettelmann, N., P\"{u}stow, R.~\& Redmer, R.~(2013). Saturn layered structure and homogeneous evolution models with different EOSs. {\it Icarus} 225, 548.\\
     \sitem[] Pollack, J.~B., Hubickyj, O., Bodenheimer, P., Lissauer, J.~J., Podolak, M.~\& Greenzweig, Y.~(1996). Formation of the Giant Planets by Concurrent Accretion of Solids and Gas. {\it Icarus}, 124, 62.\\
     \sitem[] P\"{u}stow, R., Nettelmann, N., Lorenzen, W., \& Redmer, R.~(2016). H/He demixing and the cooling behavior of Saturn. {\it Icarus}, 267, 323.\\
     \sitem[]  Riddle, A.C., Warwick, J.W.~(1976). Redefinition of system III longitude. {\it Icarus}, 27, 457--459. \\
     \sitem[] Read, P. L., Dowling, T.~E.~\& Schubert, G.~(2009). Saturn's rotation period from its atmospheric planetary-wave con- figuration. {\it Nature} 460, 608--610. \\
     \sitem[] Roberts, P.~H.~\& King, E.~(2013). On the genesis of the Earth's magnetism. {\it Rep.~Prog.~Phys.} 76, 096801 (55pp). \\
     \sitem[] Rosenblum, E., Garaud, P., Traxler, A.~\& Stellmach, S.~(2011). Erratum: "Turbulent Mixing and Layer Formation in Double-diffusive Convection: Three-dimensional Numerical Simulations and Theory". {\it ApJ}, 742, 132\\
     \sitem[] Roulston, M.S.~\& Stevenson, D.~J., (1995). Prediction of neon depletion in JupiterÕs atmosphere. {\it EOS}, 76, 343 (abstract)\\
     \sitem[] Salpeter, E.~E.~(1973). On Convection and Gravitational Layering in Jupiter and in Stars of Low Mass. {\it ApJL}, 181, L83. \\
     \sitem[] Saumon, D., Chabrier, G., \& van Horn, H.~M., (1995). An Equation of State for Low-Mass Stars and Giant Planets. {\it ApJS}, 99, 713. \\
     \sitem[] Saumon, D.~\& Guillot, T.~(2004). Shock compression of deuterium and the interiors of Jupiter and Saturn. {\it ApJ}, 609, 1170.\\
     \sitem[] Sch{\"o}ttler \& Redmer (2018). Ab Initio Calculation of the Miscibility Diagram for Hydrogen-Helium Mixtures. {\it PRL}, 120, Issue 11, id.115703.\\
     \sitem[] Schouten,  J.~A.,  de Kuijper,  A.~\& Michels,  J.~P.~J.~(1991). Critical line of He-H$_2$ up to 2500 K and the influence of attraction on fluid-fluid separation.  {\it Phys. Rev. B}, 44, 6630.\\
     \sitem[] Spilker, L.~J.~(2012). Cassini: Science highlights from the equinox and solstice missions. In: Lunar and Planetary Institute Science Conference Abstracts, 43, p. 1358.\\
     \sitem[] Stevenson, D.~J.~\& Salpeter E.~E.~(1977a). The dynamics and helium distribution in hydrogen-helium fluid planets. {\it ApJS}, 35, 239.\\
     \sitem[] Stevenson, D.~J.~\& Salpeter, E.~E.~(1977b). The phase diagram and transport properties for hydrogen-helium fluid planets. {\it ApJS}, 35, 221.\\
     \sitem[] Stevenson, D. J.~(1985). Cosmochemistry and structure of the giant planets and their satellites. {\it Icarus}, 62, 4--15.\\
     \sitem[] Vazan, A., Helled, R., Kovetz, A.~\& Podolak, M.~(2015). Convection and Mixing in Giant Planet Evolution. {\it ApJ}, 803, 32.\\
     \sitem[] Vazan, A., Helled, R., Podolak, M.~\& Kovetz, A.~(2016).  The Evolution and Internal Structure of Jupiter and Saturn with Compositional Gradients. {\it ApJ}, 829, 118.\\
     \sitem[] Vazan, A., Helled, R.~\& Guillot, T.~(2018). JupiterÕs evolution with primordial composition gradients, {\it A\&A},  610, id.L14, 5 pp. \\
     \sitem[] Venturini, J., Alibert, Y., \& Benz, W.~(2016). {\it A\&A}, 596, id.A90, 14 pp.\\
     \sitem[] von Zahn, U., Hunten, D.~M.~\& Lehmacher, G..~(1998). Helium in JupiterÕs atmosphere: Results from the Galileo probe helium interferometer experiment. {\it JGR}, 103, 22815.\\
     \sitem[] Wahl, S.~M., Hubbard,  W.~B.,  Militzer, B.~et al.~(2017) 
     Comparing Jupiter interior structure models to Juno gravity measurements and the role of a dilute core. {\it GRL}, 44, 4649--4659.\\
     \sitem[] Wisdom, J.~\& Hubbard, W.~B.~(2016). Differential rotation in Jupiter: A comparison of methods. {\it Icarus}, 267, 315.\\
     \sitem[] Wilson, H.~F.~\& Militzer, B.~(2010). Sequestration of noble gases in giant planet interiors. {\it PRL}, 104,  121101.\\
     \sitem[] Wilson, H.~F.~\& Militzer, B.~(2012). Solubility of water ice in metallic hydrogen: Consequences for core erosion in gas giant planets. {\it ApJ}, 745, 54.\\
     \sitem[] Wood, T. S., Garaud, P. \& Stellmach, S.~(2013). A new model for mixing by double-diffusive convection (semi-convection). II. The transport of heat and composition through layers. {\it ApJ}, 768, 157.\\
     \sitem[] Zaghoo, M. Salamat, A \&. Silvera, I.~F.~(2016). Evidence of a first- order phase transition to metallic hydrogen, {\it Phys.~Rev.~B}, 93, 155128.\\
 \sitem[] Zharkov, V.~N., \& Trubitsyn, V.~P.~(1978). Physics of Planetary Interiors (Astronomy and Astrophysics Series; Tucson, AZ: Pachart).
\end{itemize}
}

\clearpage


\renewcommand{\baselinestretch}{1.2}

\end{document}